\documentclass[english,aps,pre,nofootinbib,preprint,amssymb,showpacs,superscriptaddress,nobalancelastpage]{revtex4-1}
\usepackage[T1]{fontenc}
\usepackage[latin9]{inputenc}
\usepackage{color}
\usepackage{verbatim}
\usepackage{amsmath}
\usepackage{graphicx}
\usepackage{babel}
\usepackage{amsfonts}
\usepackage{bm}
\usepackage{fullpage}
\usepackage{xcolor}
\usepackage{subfig}
\usepackage{cancel}

\graphicspath{ {./obrazki/} }
\newcommand{\dd}{\textrm{d}}

\newcommand{\ddbarr}{\hspace{0.1em} \textrm{d} \hspace*{-0.2em}\bar{}\hspace*{0.02em} \hspace{0.1em}}

\begin{document}
 
\title{Direction of spontaneous processes in non-equilibrium systems with movable/permeable internal walls}

\author{Robert Ho\l yst}
\email{rholyst@ichf.edu.pl}
\affiliation{Institute of Physical Chemistry, Polish Academy of Sciences Kasprzaka
44/52, 01-224 Warszawa}

\author{Pawe\l{} J. \.{Z}uk}
\email{pzuk@ichf.edu.pl}
\affiliation{Institute of Physical Chemistry, Polish Academy of Sciences Kasprzaka
44/52, 01-224 Warszawa}

\author{Anna Macio\l ek}
\email{amaciolek@ichf.edu.pl}
\affiliation{Institute of Physical Chemistry, Polish Academy of Sciences Kasprzaka
44/52, 01-224 Warszawa}
\affiliation{Max-Planck-Institut f{\"u}r Intelligente Systeme Stuttgart, Heisenbergstr.~3,
D-70569 Stuttgart, Germany}

\author{Karol Makuch}
\email{kmakuch@ichf.edu.pl}
 \affiliation{Institute of Physical Chemistry, Polish Academy of Sciences Kasprzaka
 44/52, 01-224 Warszawa}

\author{Konrad Gi\.zy\'nski}
 \affiliation{Institute of Physical Chemistry, Polish Academy of Sciences Kasprzaka
 44/52, 01-224 Warszawa}

\date{\today}

\begin{abstract}

The second law of equilibrium thermodynamics explains the direction of spontaneous processes in a system after removing internal constraints. When the system only exchanges energy with the environment as heat, the second law states that spontaneous processes at constant temperature satisfy the following inequality: $\dd U - \ddbarr Q \leq 0$. Here, $\dd U$ is the infinitesimal change of the internal energy, and $\ddbarr Q$ is the infinitesimal heat exchanged during the process. We will consider three different systems in a heat flow: ideal gas, van der Waals gas, and a binary mixture of ideal gases. 
We will also study ideal gas and van der Waals gas in the heat flow and gravitational field.
We will divide each system internally into two subsystems by a movable wall. We will show that the direction of the motion of the wall, after release, at constant boundary conditions is determined by the same inequality as in equilibrium thermodynamics. The only difference between equilibrium and non-equilibrium law is the dependence of the net heat change, $\ddbarr Q$, on the state parameters of the system. 
We will also consider a thick wall permeable to gas particles and derive Archimedes' principle in the heat flow. Finally, we will consider the ideal gas's Couette (shear) flow. In this system, the direction of the motion of the internal wall follows from the inequality $\dd E - \ddbarr Q - \ddbarr W_s \leq 0$, where $\dd E$ is the infinitesimal change of the total energy (internal plus kinetic) and $\ddbarr W_s$ is the infinitesimal work exchanged with the environment due to shear force imposed on the flowing gas. Ultimately, we will synthesize all these cases in a general framework of the second law of non-equilibrium thermodynamics.

\end{abstract}
\maketitle
\section{Introduction} \label{sec:intr}

The direction of spontaneous processes follows the second law of equilibrium thermodynamics \cite{callen1998thermodynamics}. 
It tells us that isolated systems reaching equilibrium increase their entropy. The second law is practical. It allows us to predict which chemical reactions occur spontaneously, at what thermodynamic conditions liquids evaporate or freeze, and how to design efficient engines.
Such a law, although needed, has yet to be discovered for out-of-equilibrium systems characterized by the continuous flux of energy flowing across them.
This remains true despite large efforts \citep{Introduction_to_thermodynamics_of_irreversible_processes_Ilya_Prigogine,oono1998steady,sekimoto1998langevin,hatano2001steady,sasa2006steady,guarnieri2020non,boksenbojm2011heat,netz2020approach,mandal2013nonequilibrium,holyst2019flux,jona2014thermodynamics,speck2005integral,mandal2016analysis,maes2019nonequilibrium,Zhang2021,glansdorff1964general,maes2014nonequilibrium,ruelle2003extending,sasa2014possible,komatsu2008expression,komatsu2008steady,komatsu2011entropy,chiba2016numerical,nakagawa2017liquid,nakagawa2019global,sasa2021stochastic,nakagawa2022unique},
and sucesses are limited either to the isothermal
situations or to the small temperature differences \citep{komatsu2008expression,komatsu2008steady,komatsu2011entropy,chiba2016numerical,nakagawa2017liquid,nakagawa2019global,sasa2021stochastic,nakagawa2022unique}.
Promising results were obtained in small scale systems allowing for a discrete description \cite{VANDENBROECK20156,Rao2018}, and they pave a trajectory approach to macroscopic stochastic thermodynamics \cite{falasco2024macroscopic}. 
The non-equilibrium thermodynamics today (called thermodynamics of irreversible processes) is the set of non-linear differential equations representing the conservation of mass, momentum, and energy~\cite{deGroot2013,jou1996extended}.
The second law is missing in this description. Various attempts to present the widely applicable formalism of the second law based on these equations failed \cite{PhysRevE.50.1645}.
Although most of the world around us is not at equilibrium, we do not have the general tools to predict the direction of spontaneous processes occurring within them. This paper will show preliminary observations for many different systems, which can guide us towards formulating the second law of non-equilibrium thermodynamics. 

Non-equilibrium states involve macroscopic energy fluxes flowing across the system \cite{deGroot2013}. These fluxes are sustained within the system by the non-vanishing gradients of temperature (in the case of heat flow), pressure (in the case of mass flow), or chemical potential (in the case of diffusion of particles).
Thus, a non-equilibrium state has non-uniform temperature, pressure, chemical potential, and velocity. The local equations of state relating internal energy and pressure to density and temperature and spatial profiles of density, temperature, and concentrations fully characterize non-equilibrium states.
In our recent papers, we formulated the first law of global thermodynamics for various non-equilibrium systems \cite{Makuch2022,Holyst2023,e25111505} including, for example, gravity \cite{Holyst2023} or Couette flow \cite{Karol:JCP2023}. 
We represented internal energy as a function of a few global state parameters. 
These parameters were obtained by mapping a non-equilibrium and, by definition, non-uniform system into a uniform one. 
We averaged local equations of state over the system's volume. 
The averaging resulted in global equations of state, which we wrote in the same form as at equilibrium. These global non-equilibrium equations of state included new state parameters. For example, the internal energy of the van der Waals gas subjected to continuous heat flux $U(S^*, V, N, a^*,b^*)$ is a function of 5 parameters of state, where $S^*$ is the non-equilibrium entropy, $V$ is the volume, $N$ is the number of particles and $a^*$ and $b^*$, new parameters of state, are renormalized van der Waals interaction parameters. The net heat that flows in/out of the system to change the internal energy of the van der Waals gas is given by:
\begin{equation}
	\label{eq:1}
	\ddbarr Q=T^{*} \dd S^{*}-\frac{N^{2}}{V} \dd a^{*}+Nk_{B}T^{*}\left(\frac{V}{N}-b^{*}\right)^{-1} \dd b^{*}.
\end{equation}
The $a^*$ and $b^*$ state parameters appear in the net heat differential because, in a non-uniform system, the change of the density profile leads to the local absorption or release of heat. In general, under our construction of global thermodynamics, all material parameters in the equilibrium equations of states become parameters of state of the non-equilibrium state.

In a system kept at a constant temperature at equilibrium, the energy is exchanged with the environment as heat only.
The second law of equilibrium thermodynamics states \cite{callen1998thermodynamics,Holyst2012} that the Helmholtz free energy, $F(T, V, N, x)$, is minimized for $x$ (the variable describing internal constraint) at constant $T, V, N$ (temperature, volume, and number of particles, respectively). 
The minimization defines the equilibrium value of $x$. The change of $F$ as a function of $x$ is negative if the initial $x$ does not correspond to the equilibrium state. Thus, the change of $F$ when we move the system from initial to final $x$ is given by $\Delta F\le 0$.
Rewriting this equation using internal energy and entropy gives $\Delta U-T\Delta S\le 0$. In the infinitesimal form we get $\dd U-T \dd S \le 0$. 
Finally, the net heat (heat that enters or leaves the system and changes the internal energy) is $\ddbarr Q = T \dd S$. In general, the second law states $\dd U-\ddbarr Q\le 0$. Thus, the system minimizes part of its internal energy. This part does not account for the energy, which, at equilibrium, is continuously exchanged with the environment (here, net heat). 

In this paper, we elucidate non-equilibrium systems in a continuous heat flow. 
Out of equilibrium, net heat becomes the amount of energy that enters or leaves the system in the form of heat and changes the internal energy.
We show the consequences of applying
 $\dd U- \ddbarr Q\le 0$ in the systems' stationary (steady) states. 
In section II, we discuss the ideal gas, van der Walls gas and the binary mixture of ideal gases enclosed by two fixed walls of different temperatures.
We calculate $\dd U$ and $\ddbarr Q$ as a function of state parameters. 
From (\ref{eq:1}), we see that $\ddbarr Q$ differs from what we know from equilibrium. 
Similarly, as is done at equilibrium, we divide the enclosed gas with an internal diathermic wall and check that $\dd U-\ddbarr Q\le 0$ determines the spontaneous motion of the wall towards the stationary position. 
In section III, we extend the problem to account for gravity and heat flow in the case of ideal gas and van der Waals gas. Next, we assume that the wall has a final thickness and is permeable (porous). 
That leads to the derivation of Archimedes' principle from the second law of thermodynamics.
Section IV considers the ideal gas heated volumetrically and discusses the transition between two locally stable states to compare their stability.
In section V we study the ideal gas in the shear flow (Couette flow) and generalize the second law to the case when, apart from heat, the system also exchanges energy in the form of work.
The work originates from the shear forces imposed on the gas flow.
We finish the paper by generalizing the second law in the discussion section.

\section{The ideal gas, gas mixtures and van der Waals gas in the heat flow} \label{sec:2}

\begin{figure}[htb!]
 \includegraphics[width=0.5\textwidth]{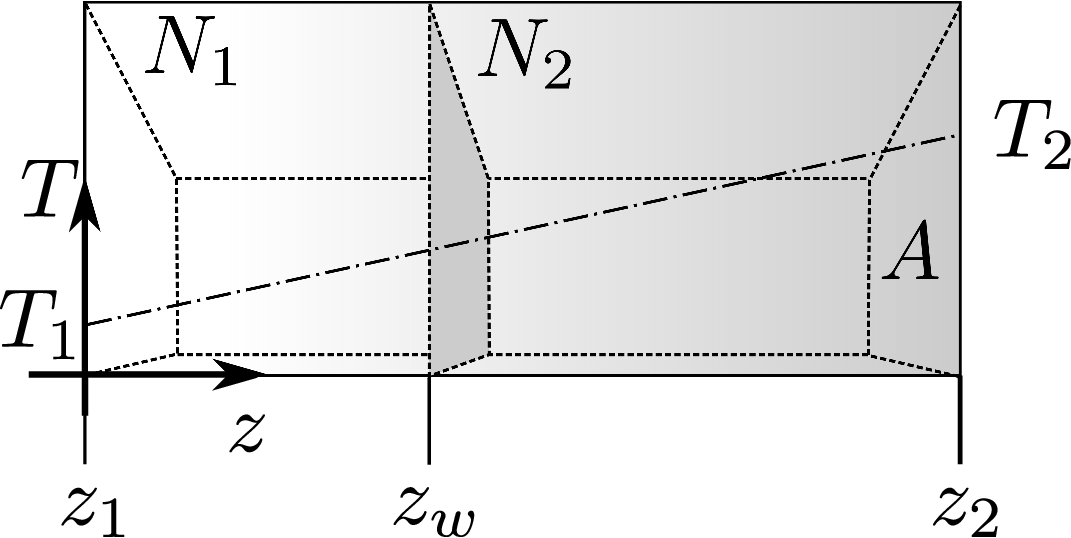}
 \caption[option]{Schematic of a gas confined between two parallel walls of area $\mathrm{A}$.
The system is assumed to be translationally invariant in the $x$ and $y$ directions.
The thin interior wall at $z_w$ is treated as an internal constraint.
We assume that the internal wall is diathermic, impenetrable and can move freely. 
The inner wall divides the system into two compartments containing $N_1$ and $N_2$ gas particles ($N_1+N_2=N$).
The outer wall placed at $z_1=0$ has a fixed temperature $T_1$, and the outer wall placed at $z_2=L$ is kept at a fixed temperature $T_2 >T_1$.
The resulting linear temperature profile is indicated by the dash-dot line.
 }
\label{fig:fig1}
\end{figure}

{\bf Ideal gas.} The model geometry is spanned between two parallel walls at $z_1=0$ and $z_2=L$ as shown in  Fig.~\ref{fig:fig1}.
For now, let's focus on the arrangement that does not include the inner wall.

The area of each wall $A$ is large enough so that the system is translationally invariant in $x,y$ directions.
The walls are in contact with thermostats,
which are maintained at different temperatures $T_1 < T_2$. 
In the presence of heat flux, quantities specifying equilibrium thermodynamic states, such as number density $n=N/V$, pressure $p$ or temperature $T$, become space-dependent.
The profiles $n(z)$, $p(z)$, and $T(z)$ of these quantities can be determined using, for example,  the irreversible thermo-hydrodynamics approach,
which rests on local equilibrium assumption and represents the conservation of mass, momentum and energy,  supplemented with the equation of states and relations between fluxes and thermodynamic forces.
For a stationary state with a vanishing velocity field, the thermo-hydrodynamics problem reduces to the 
 constant pressure condition $p\left(z\right)=p$.

The construction of global thermodynamics describing the non-equilibrium steady state requires mapping it to a uniform system characterized by a finite number of state parameters.
Mapping involves averaging the local pressure and energy over the system's volume so that, when averaged, the global equations of state have the same form as at equilibrium.
For an ideal gas with $f$ degrees of freedom, this procedure gives for a steady state with no mass flow
\begin{equation} \label{eq:2.1}
\begin{aligned}
& p=\frac{1}{L}\int_{0}^{L}\,n\left(z\right)k_{B}T\left(z\right) dz= nk_{B}T^{*}, \\
& U=\frac{f}{2}A\int_{0}^{L}\,n\left(z\right)k_{B}T\left(z\right) dz= \frac{f}{2}Vnk_{B}T^{*},
\end{aligned}
\end{equation}
where $n=N/V=(1/L)\int_0^L n(z)dz$ is  the average number density and 
$T^{*}$ is given by 
\begin{equation} \label{eq:2.2}
T^{*}=\frac{A\int_{0}^{L}dz\,n\left(z\right)T\left(z\right)}{A\int_{0}^{L}dz\,n\left(z\right)}.
\end{equation}
The new state parameter $T^{*}$ can be interpreted as the average temperature of the uniform system onto which we have mapped the original, non-equilibrium one.
The explicit expression for $T^{*}$ depends on the specific temperature and number density profile.
In the case of constant pressure~\cite{Makuch2022}, 
the equations of irreversible thermo-hydrodynamics
give  $T(z)= T_1+(T_2-T_1)(z/L)$,  $n(z)=  n\left ( T_2-T_1\right)/\left[\log (T_2/T_1)T(z)\right]$
and as a result
\begin{equation} \label{eq:2.2a}
 T^* =(T_2-T_1)/\ln (T_2/T_1).
\end{equation}
Because  mapping gives us the same formal structure as we know from equilibrium, the variable  $S^*$  conjugate to $T^{*}$ is given by 
\begin{equation} \label{eq:2.3}
 S^{*}\left(U,V,N\right)/k_B=N\log\left[\left(\frac{U}{(f/2)k_BN}\right)^{f/2}\frac{V}{N}\right]+Ns_{0}/k_{B},
\end{equation}
where  the constant $s_{0}$  is taken such  that $S^{*}$
for $T_{2}=T_{1}$ gives the equilibrium expression for the entropy~\citep{callen1998thermodynamics}.
The  internal energy of a non-equilibrium steady state is thus a function of three parameters of state $U\left(S^{*},V,N\right)$, with
the thermodynamic relation:
\begin{align}
\left(\frac{\partial S^{*}}{\partial U}\right)_{V,N} & =\frac{1}{T^{*}},\label{eq:2.4}\\
\left(\frac{\partial S^{*}}{\partial V}\right)_{U,N} & =\frac{p}{T^{*}}.\nonumber 
\end{align}

{\bf Binary mixture.} Mixing two ideal gases $a$ and $b$ with $f_a$ and $f_b$ degrees of freedom, respectively, add to the temperature and density profile the profile of concentration induced by a heat flux.
As a result, the mapping procedure leads to a uniform system with two new state parameters. 
In  addition to $T^{*}$ we have  two effective degrees of freedom   $f_a^*$ and 
and $f_b^*$, which, however, are not independent. They  allow us to write the internal energy and pressure in the same form as in equilibrium:
\begin{equation}
\begin{aligned}\label{eq:2.5}
& p = n_ak_BT^* + n_bk_BT^*, \\
& U  =\frac{f_a^*}{2}Vn_ak_B T^* + \frac{f_b^*}{2}Vn_bk_B T^*,
  \end{aligned}
\end{equation} 
where $n_a=N_a/V$ and $n_b=N_b/V$ are the average number densities of the first and second component of the mixture, respectively. 
The effective degrees of freedom are obtained as 
\begin{equation}
\begin{aligned}
\label{eq:2.6}
  f_i^*= \frac{f_i}{x_i}\frac{1}{L}\int_0^L x_i(z)dz, \quad i=a,b
 \end{aligned}
\end{equation} 
where $x_i=n_i/n$ is the average number fraction of $i$ component and $x_i(z)$ is its profile; $x_a+x_b=1$ in the absence of chemical reactions.
$f_a^*$ and $ f_b^*$  are related via 
 $\left(f_a^*/f_a\right)x_a+\left(f_b^*/f_b\right)x_b=1$.
The non-equilibrium entropy $S^{*}$ has the same form as in equilibrium but with $T$ replaced by $T^*$ and $f_i$ by $f_i^*$. 
It is the sum of the entropy of the two components of the mixture considered separately and the entropy of mixing:
\begin{equation} \label{eq:2.7}
\begin{aligned}
S^*\left(U,V,N_a,N_b,f_a^*\right)/k_B &= N_a\left[\frac{f_a^*}{2}+1+\ln \frac{V}{N_a}\left(\frac{2U}{k_B(N_af_a^*+N_bf_b^*)}\right)^{f_a^*/2} \right]\\
&+ N_b\left[\frac{f_b^*}{2}+1+\ln\frac{V}{N_b}\left(\frac{2U}{k_B(N_af_1^*+N_bf_2^*)}\right)^{f_b^*/2}\right] \\
&-N_a\ln \frac{N_a}{N}-N_b\ln \frac{N_b}{N}+S_0/k_B.
\end{aligned}
\end{equation}
Here, $S_0$ may depend on $N_a$ and $N_b$. Again, we choose it such that $S^*$ reduces to the equilibrium entropy of a binary mixture in the absence of heat flux.
The internal energy of a binary mixture of ideal gases in the non-equilibrium steady state is thus a function of five parameters of state $U\left(S^{*},V,N_a,N_b,f_a^*\right)$.

{\bf Van der Waals gas.} Gas of interacting particles obeying the van der Waals equations of state is described with two additional interaction parameters $a$ and $b$.
As a result of the mapping procedure, we obtain
\begin{equation}\label{eq:2.8}
\begin{aligned}
&p=\frac{nk_{B}T^{*}}{1-nb^*}-a^{*}n^{2},\\
&U=\frac{f}{2}Vnk_{B}T^{*}-a^{*}Vn^{2}.
\end{aligned}
\end{equation}
Three new state parameters describe the van der Waals gas in the steady state.
Beside the effective temperature $T^*$  defined  by the same expression (\ref{eq:2.2}) as for ideal gas, 
we have the effective interaction parameter $a^*$ given by 
\begin{equation}\label{eq:2.9}
 a^{*} = \frac{1}{L} \int_{0}^{L} \frac{a \,n\left(z\right)^{2}}{n^{2}} dz,
\end{equation}
and $b^*$ defined by the formula
\begin{equation}\label{eq:2.10}
 \frac{nk_{B}T^{*}}{1-nb^*} = \frac{1}{L}\int_{0}^{L} \frac{a\,n\left(z\right)k_BT(z)}{1-bn(z)} dz.
\end{equation}
Because Eqs.~(\ref{eq:2.8}) have the same structure as in equilibrium, the non-equilibrium entropy $S^{*}$ has the same form as in equilibrium but with $T$ replaced by $T^*$,  $a$ by $a^*$ and $b$ by $b^*$. 
\begin{equation}
 \label{eq:2.11}
 S^{*}\left(U,V,N,a^{*},b^{*}\right)/k_B=N\log\left[\left(\frac{U+\frac{a^*N^2}{V}}{(f/2)k_BN}\right)^{f/2}\frac{V-Nb^*}{N}\right]+Ns_{0}/k_{B}.
\end{equation}
The internal energy of van der Waals gas in the non-equilibrium steady state is thus a function of five parameters of state $U\left(S^{*},V,N,a^{*},b^{*}\right)$.

Note that for all systems discussed above, the non-equilibrium entropy $S^*$ is part of the total entropy $S_{tot}$ defined as the volume entropy density $s(z)$ integral over the system volume.
It contains information about the heat absorbed/released in the system on top of the dissipative background (temperature profile).

{\bf Net heat.} In the case of a very slow transition between stationary states  by, e.g., a slight change of temperature $T_2$ or by the change of a distance $L$ between  the confining wall $L$, the energy changes only by means of mechanical work and heat flow: 
\begin{equation} \label{eq:2.12}
 \dd U= \ddbarr Q + \ddbarr W, \qquad \textrm{with} \qquad \ddbarr W = - p \dd V,
\end{equation}
where $\ddbarr Q$ is the net heat transferred to the system during a small change between two non-equilibrium stationary states.
Thus, the above equation can be considered as the first law of non-equilibrium thermodynamics.
Using the fundamental relations for each of the above examples (Eqs.~(\ref{eq:2.3}), (\ref{eq:2.7}), and (\ref{eq:2.11})), we can determine the net heat from the energy balance $\dd U=\ddbarr Q - p \dd V$.
For an ideal gas, we have
\begin{equation} \label{eq:2.13}
 \dd U=T^* \dd S^*- p \dd V,   \qquad  \textrm{and} \qquad \ddbarr Q =T^* \dd S^*,
\end{equation}
which has the same formal form as in equilibrium.

This is different for a binary mixture of ideal gases, where the net heat acquires additional terms. 
In general 
\begin{equation}
\label{eq:2.14}
\dd U=T^* \dd S^*-p \dd V +\mu_a^* \dd N_a+\mu_b^* \dd N_b+\mathcal{F}_a \dd f_a^*+\mathcal{F}_b \dd f_b^*,
\end{equation}
and for fixed number of components $N_a$ and $N_b$, we have
\begin{equation}
\label{eq:2.15}
 \ddbarr Q=T^* \dd S^*+\mathcal{F}_a \dd f_a^*+\mathcal{F}_b \dd f_b^*,
\end{equation}
where  $\dd f_b^* = -\frac{x_bf_a}{x_af_b} \dd f_a^*$  and  $\frac{\mathcal{F}_i}{T^*}=-\left(\frac{\partial S^*}{\partial f_i^*}\right)_{U,V,N_a,N_b,f_{j\ne i}^*}$ with $S^*$ given by Eq.~(\ref{eq:2.7}). 

Additional terms in the net heat also occur for the van der Waals gas, where
\begin{equation}\label{eq:2.16}
\dd U=T^{*} \dd S^{*}-p \dd V-\frac{N^{2}}{V}\dd a^{*}+nk_BT^*\left(\frac{V}{N}-b^*\right)^{-1} \dd b^*
\end{equation}
so that
\begin{equation}\label{eq:2.17}
\ddbarr Q=T^{*} \dd S^{*}-\frac{N^{2}}{V} \dd a^{*}+nk_BT^*\left(\frac{V}{N}-b^*\right)^{-1} \dd b^*
\end{equation}

{\bf Movable internal wall.} We introduce to the considered systems an internal constraint in the form of a wall parallel to the bounding walls (Fig.~\ref{fig:fig1}). 
We assume the wall is thin, freely movable, diathermal and impenetrable.
The internal constraint divides the system into two subsystems, 1 and 2, each with a fixed number of particles $N_1$ and $N_2$, and the volumes $V_1$ and $V_2$. We choose the volume of one subsystem $V_1$ as the parameter representing constraint.
The total energy is the sum of the energies of the two subsystems: 
\begin{equation}
 \label{eq:2.28}
 U(S_1^*,V_1,N_1,S_2^*,V-V_1,N_2,\ldots)= U_1(S_1^*,V_1,N_1,\ldots)+U_2(S_2^*,V-V_1,N_2,\ldots),
\end{equation}
where $\ldots$ denotes possible other new state variable, such as $f_a^*$ for the binary mixture of ideal gases or $a^*$ and $b^*$ for the van der Waals gas. 
The steady position of the wall, which moves without friction, is determined by the equality of the pressures exerted by each subsystem $p_1 = p_2$.
The effective temperatures of both subsystems are~\cite{Makuch2022}
\begin{align*}
T_{1}^{*} & =\frac{\frac{V_{1}}{V}\left(T_{2}-T_{1}\right)}{\log\left(\frac{T_{1}+\frac{V_{1}}{V}\left(T_{2}-T_{1}\right)}{T_{1}}\right)},\\
T_{2}^{*} & =\frac{\left(1-\frac{V_{1}}{V}\right)\left(T_{2}-T_{1}\right)}{\log\left[\frac{T_{2}}{T_{1}+\frac{V_{1}}{V}\left(T_{2}-T_{1}\right)}\right]}.
\end{align*}
Using the explicit forms of $\dd U_1, \dd U_2$ and $\ddbarr Q_1, \ddbarr Q_2$, we can see that  this condition can be obtained from the following minimum principle
\begin{equation} \label{eq:2.19}
 \dd U_1 + \dd U_2 - \ddbarr Q_1 - \ddbarr Q_1 = -\left(p_1 - p_2   \right)  \dd V_1 \leq 0 \; ,
\end{equation}
which we call the second law of non-equilibrium thermodynamics.
Thus, the difference between the total energy of the system and the heat exchanged with the environment is minimized while the internal wall moves to the new position.  
The equality defines the condition of the stationary position of the wall,
given by the equality of pressures $p_1 = p_2$ in two subsystems.
For the specific case of an ideal gas, we have 
\begin{equation}\label{eq:2.20}
 p_1=k_B\frac{N_1}{V}\frac{T_2-T_1}{\ln \left[\frac{T_1+(T_2-T_1)\frac{V_1}{V}}{T_1}\right]} \qquad \qquad p_2=k_B\frac{N_2}{V}\frac{T_2-T_1}{\ln \left[\frac{T_2}{T_1+(T_2-T_1)\frac{V_1}{V}}\right]}
\end{equation}
Demanding $p_1=p_2$ and solving for $V_1$ gives us the position of the internal wall at the stationary state for given temperatures $T_2$ and $T_1$:
\begin{equation}
\label{eq:2.21}
\frac{V_1}{V}=\frac{\left(\frac{T_2}{T_1}\right)^{N_2/N}-1}{\frac{T_2}{T_1}-1}.
 \end{equation}
We note that for the diathermal internal wall, there is only a single solution for $V_1$ for all possible values of $T_2>T_1$.
In Sec.~\ref{sec:4}, we consider an adiabatic internal wall with volumetric heating and show that, in this case, there might be more solutions for $V_1$, among which we can compare stability.

\section{Gravitational field: ideal gas, van der Waals gas and Archimedes principle} \label{sec:3}

\begin{figure}[!hbt]
\begin{center}
\includegraphics[width=\textwidth]{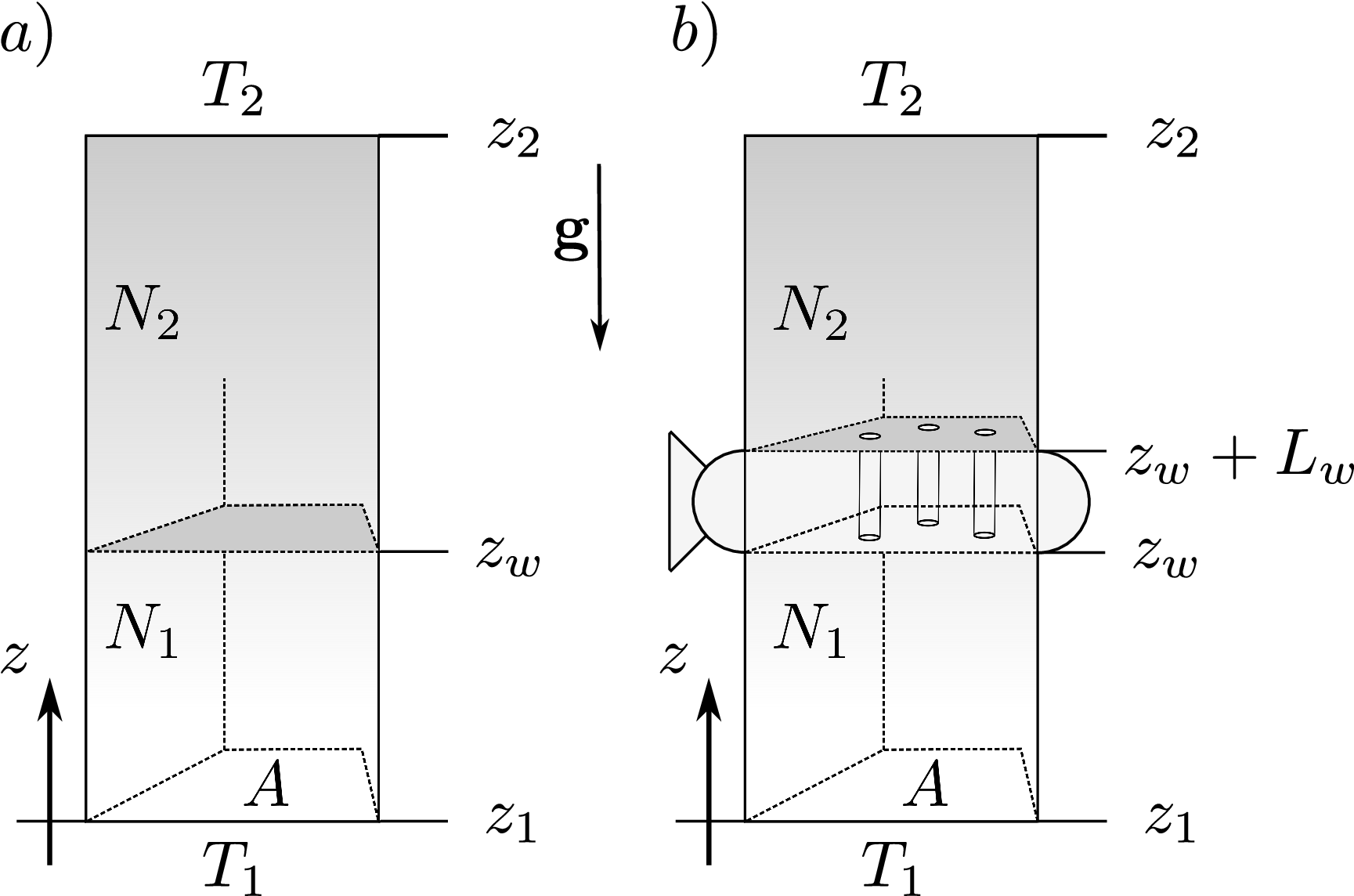} 
\end{center}
\caption{Gas column in gravity.
A column of height $L$ contains $N$ moles of gas particles.
The base ($z_{1}$) and top ($z_{2}$) are in contact with temperature reservoirs kept at two different temperatures: $T_{1}$ at the base and $T_{2}$ at the top.
The column's base and top surface area equal $A$, and the system is translationally invariant in $x,y$ directions.
a) Impermeable and diathermal wall at $z_w$ separates bottom and top sections with $N_1$ and $N_2$ particles of gas ($N_1+N_2=N$).
b) Permeable, thick ($L_w$), diathermal wall at $z_w$ separates sections containing $N_1$ and $N_2$ particles of gas ($N_1+N_2=N$).}
\label{fig:columnWithWall} 
\end{figure}

{\bf Ideal gas in gravity.}  We insert the gas column into constant, external gravitational field $-\hat{\textbf{e}}_z g$ that acts in the direction of the heat flux (see Fig.~\ref{fig:columnWithWall}). 
Here, the situation of interest is without macroscopic fluid motion, which means there is no Rayleigh-Bennard convection \cite{Chandrasekhar2013,Getling1998,Mizerski2021}.
The meanings of symbols from previous sections remain valid. 
The non-equilibrium entropy of a single gas column of an ideal gas is \cite{Holyst2023}
\begin{equation}
 S^* = \frac{3}{2} N k_B \ln \frac{U - \frac{N M^* g L }{2} - N M g z_{1} }{ U_0 - \frac{ N M_0 g L_0 }{2} -  N M g z_{1,0}  } + N k_B \ln \frac{V}{V_0} + S_0,
 \label{eq:entropyGrav}
\end{equation}
where $L=z_2 - z_1$ is the height of the gas column, $z_1$ is the coordinate of the column base, $z_2$ is the coordinate of the column top, $V=L A$ is the volume of the column with $A$ being surface area of the column base,  and $M$ denotes molecular mass of the gas.
The lower index $0$ indicates reference values.
In the above, state parameter $M^{*}$ is renormalized mass defined with the help of the potential gravitational energy of the gas \cite{Holyst2023}
\begin{equation}
E_{\textrm{pot}}=\frac{NM^{*}gL}{2} 
\Leftrightarrow 
M^{*} = \frac{2 E_{\textrm{pot}}}{N g L},
\end{equation}
which is contained inside the column
\begin{equation}
E_{\textrm{pot}} = 
gAL^{2}\int_{z_{1}/L}^{z_{2}/L}\rho z'\dd z'.
\end{equation}
The state parameter $M^{*}$ is coupled to the gravitational field and informs about the system's potential energy with respect to the position of the column base.
To include all effects of the gravity,
we need to account also for the potential energy of the base of the column
\begin{equation}
 E_{\textrm{pot},0} = N M g z_1.
\end{equation}
The total internal energy of the gas column
\begin{subequations}
\label{eq:perfGasState}
\begin{equation}
U=\frac{3}{2}Nk_{B}T^{*}+\frac{NM^{*}gL}{2} + N M g z_1,
\label{eq:eq state U}
\end{equation}
consists of thermal and gravitational contributions.
Thermal contribution (through mapping procedure) defines renormalized temperature $T^{*}$
\begin{equation}
 E_{\textrm{T}} 
 = \frac{3}{2} N k_B T^{*} 
 = AL\int_{z_{1}/L}^{z_{2}/L}\frac{3k_{B}}{2M}\rho T\dd z'
 \Leftrightarrow
 T^{*} = \frac{V}{N M} \int_{z_{1}/L}^{z_{2}/L} \rho T \dd z'.
\end{equation}
The average pressure inside the column is just like in the case without the gravity (\ref{eq:2.1})
\begin{equation}
p_{\textrm{av}}=\frac{Nk_{B}T^{*}}{V}.
\label{eq:eq state pav}
\end{equation}
It must be so, as changing the surface area of the column base does not change the gravitational energy, and all work is done against the thermal motion of gas particles.
However, the presence of an external field introduces the anisotropy in the pressure profile.
Pressures at the top
\begin{equation}
p(z_{2})=\frac{Nk_{B}T^{*}}{V} -\frac{NM^{*}gL}{2 V},\label{eq:eqstate pzL}
\end{equation}
and bottom
\begin{equation}
p(z_{1})=p(z_{2}) + \frac{NMgL}{V}. \label{eq:eqstate pz0}
\end{equation}
\end{subequations}
of the column are different.
Presented formulas are consistent with the functional form of $S^*$ (\ref{eq:entropyGrav}) and the internal energy differential
\begin{widetext}
 \begin{align} \label{eq:energyBalanceGravConsMass}
 \dd U 
 = & T^{*} \dd S^{*} + N\frac{gL}{2} \dd M^{*} 
 - \left(\frac{N k_{B}T^{*}}{V}-\frac{N M^{*}gL}{2V}\right)\frac{V}{L} \dd L - \frac{Nk_{B}T^{*}}{V}\dd V + N M g \dd z_1
 \nonumber \\
 = & T^{*} \dd S^{*} + N\frac{gL}{2} \dd M^{*} + \ddbarr W
 = \ddbarr Q + \ddbarr W.
 \end{align}
\end{widetext}
In the above, we distinguish two separate ways of changing the internal energy. First is through heat
\begin{subequations}
\begin{equation}
 \ddbarr Q = T^{*} \dd S^{*} + N\frac{gL}{2} \dd M^{*},
\end{equation}
which includes entropic component and gravitational interactions between the external field and the mass inside the column. 
The second is through mechanical interactions exerted on the system boundaries by the means of mechanical work
\begin{equation}
\ddbarr W = - \left(\frac{N k_{B}T^{*}}{V}-\frac{N M^{*}gL}{2V}\right)\frac{V}{L} \dd L - \frac{Nk_{B}T^{*}}{V}\dd V + N M g \dd z_1.
\end{equation}
\end{subequations}
We deliberately wrote the work differential to distinguish column elongation, changes in the column volume and changes in the column base position. Column elongation is coupled to the pressure at its top, and volume changes to the average pressure.

In the two-compartment system (Fig.~\ref{fig:columnWithWall}a)), inside the bottom part, there are $N_1$ particles and inside the top part, $N_2$ particles of an ideal gas.
We separate compartments with a thin, diathermal wall at $z=z_w$.
We assume there are no changes in column cross-section area $\dd A = 0$ and set this contribution to energy differential equal 0.
We rewrite the energy balances for the bottom $U_1$ and top $U_2$ segments explicitly to take into account the displacement of the separating wall
\begin{widetext}
 \begin{subequations}
  \begin{align}
 \dd U_1 
 = & T^{*}_1 \dd S^{*}_1 + N_1\frac{g(z_w-z_1)}{2} \dd M^{*}_1 - \left(\frac{N_1 k_{B}T^{*}_1}{V_1}-\frac{N_1 M^{*}_1 g (z_w-z_1)}{2 V_1}\right) \frac{V_1}{(z_w-z_1)} \dd z_w 
 \nonumber \\
 = & T^{*}_1 \dd S^{*}_1 + N_1\frac{g(z_w-z_1)}{2} \dd M^{*}_1 - p_1 (z_w) A \dd z_w 
 = \ddbarr Q_1 + \ddbarr W_1,
  \end{align}
  \begin{align}
 \dd U_2 
 = & T^{*}_2 \dd S^{*}_2 - N_2\frac{g(z_2-z_w)}{2} \dd M^{*}_2 + \left( \frac{N_2 k_{B}T^{*}_2}{V_2}-\frac{N_2 M^{*}_2 g (z_2-z_w)}{2 v_2}\right) \frac{ V_2 }{(z_2-z_w)} \dd z_w + N_2 M g \dd z_w
 \nonumber \\
 = & T^{*}_2 \dd S^{*}_2 + N_2\frac{g(z_2-z_w)}{2} \dd M^{*}_2 + p_2(z_2) A \dd z_w + N_2 M g \dd z_w 
 = \ddbarr Q_2 + \ddbarr W_2.
  \end{align}
 \end{subequations}
\end{widetext}
The sign in front of $p_2$ is consistent with the direction of changes $\dd z_w$. 
The second law of stationary thermodynamics states that we should take the change in internal energies and subtract all energies exchanged with the environment during that process.
For the column subject to gravity with fixed base and top, this is heat
\begin{widetext}
 \begin{align}
  \dd U_1 + \dd U_2 - \ddbarr Q_1 - \ddbarr Q_2  = \left( A \left( p_2(z_2) - p_1(z_w) \right) + N_2 M g  \right)  \dd z_w \leq 0.
 \end{align}
\end{widetext}
We find that the wall will position itself where the force necessary to expand the bottom part of the column balances the sum of the forces required to compress the top part of the column and, additionally, to raise its weight.  

{\bf Van der Waals gas in gravity}
We combine results from previous sections and construct an analogous description of the Van der Waals gas in a gravitational field. 
For the single compartment, the stationary entropy is 
\begin{equation} \label{eq:fundamentalRelationGasGravity}
 S^* = \frac{3}{2} N k_B \ln \frac{U - \frac{N M^* g L }{2} - N M g z_{1} +  a^{*} \frac{N^{2}}{V}}{ U_0 - \frac{ N_0 M_0 g L_0 }{2} -  N_0 M g z_{1,0} + a^{*}_0 \frac{N^{2}_0}{V_0}  } + N k_B \ln \frac{V - N b^{*} }{V_0 - N_0 b_0^{*}} + S_0.
\end{equation}
The internal energy of the Van der Waals gas in a gravitational field is given by
\begin{subequations}
\label{eq:vdwGasState}
\begin{equation}
U=\frac{3}{2}Nk_{B}T^{*}+\frac {NM^{*}gL}{2} + N M g z_1 - a^{*} \frac{N^{2}}{V}.\label{eq:eq state U}
\end{equation}
and pressures are
\begin{equation}
p_{\textrm{av}}= \frac{Nk_{B}T^{*}}{V-Nb^{*}} - a^{*} \frac{N^{2}}{V^2}
\end{equation}
average over the system
\begin{equation}
p(z_{2}) = \frac{Nk_{B}T^{*}}{V-Nb^{*}} - a^{*} \frac{N^{2}}{V^2} -\frac{NM^{*}gL}{2 V} ,\label{eq:eqstate pzL}
\end{equation}
at the top of the column, and
\begin{equation}
p(z_{1})=p(z_{2}) + \frac{NMgL}{V}. \label{eq:eqstate pz0}
\end{equation}
\end{subequations}
at the bottom of the column.
The energy differential has form
\begin{align} \label{eq:energyBalanceGravConsMass}
 \dd U 
 = & T^{*} \dd S^{*} + N\frac{gL}{2} \dd M^{*} 
- \left(\frac{N k_{B}T^{*}}{V-N b^{*}} - a^{*} \frac{N^2}{V^2} -\frac{N M^{*}gL}{2 V}\right)\frac{V}{L} \dd L 
  \nonumber \\
   &
 - \left( \frac{Nk_{B}T^{*}}{A \left(V- N b^{*} \right)} - a^{*} \frac{N^{2}}{V^2} \right) \dd V - \frac{N^{2}}{V} \dd a^{*}+Nk_{B}T^{*}\left(\frac{V}{N}-b^{*}\right)^{-1} \dd b^{*}
 \nonumber \\
 & + N M g \dd z_1
 \nonumber \\
 = & T^{*} \dd S^{*} + N\frac{gL}{2} \dd M^{*} - \frac{N^{2}}{V} \dd a^{*}+Nk_{B}T^{*}\left(\frac{V}{N}-b^{*}\right)^{-1} \dd b^{*} + \ddbarr W
 \nonumber \\
 & = \ddbarr Q + \ddbarr W,
 \end{align}
where we distinguish the heat differential including gravitational effects
\begin{subequations}
\begin{equation}
 \ddbarr Q = T^{*} \dd S^{*} + N\frac{gL}{2} \dd M^{*} - \frac{N^{2}}{V} \dd a^{*}+Nk_{B}T^{*}\left(\frac{V}{N}-b^{*}\right)^{-1} \dd b^{*},
\end{equation} 
and the mechanical work
\begin{equation}
\ddbarr W = 
- \left(\frac{N k_{B}T^{*}}{V-N b^{*}} - a^{*} \frac{N^2}{V^2} -\frac{N M^{*}gL}{2 V}\right)\frac{V}{L} \dd L
- \left( \frac{Nk_{B}T^{*}}{A \left(V- N b^{*} \right)} - a^{*} \frac{N^{2}}{V^2} \right) \dd V
+ N M g \dd z_1.
\end{equation}
\end{subequations}
Likewise, for the ideal gas, we write the second law of stationary thermodynamics 
\begin{widetext}
 \begin{align}
  \dd U_1 + \dd U_2 - \ddbarr Q_1 - \ddbarr Q_2 
  = \left( A \left(p_2(z_2) - p_1(z_w) \right) + N_2 M g  \right)  \dd z_w \leq 0.
 \end{align}
\end{widetext}
Again, it reads that the wall will rest where the force necessary to extend the bottom compartment matches the force necessary to compress the top compartment plus the force to lift its weight.

{\bf Archimedes principle.}
Novel phenomena appear when the wall separating compartments is permeable. 
Assume that it has a finite thickness $L_w$ with 
its own mass $M_b$ and is pierced with small channels (pores) of negligible volume that allow for the passage of gas molecules (Fig.~\ref{fig:columnWithWall}(b)). 
We parametrize the position of the wall bottom line at $z_w$.
The porous wall separating the two compartments constitutes a separate thermodynamic system.
We make additional assumptions regarding the wall: gas that fills the pores inside the wall has volume $V_3 \ll V_2,V_1$,
but each pore is large enough so that gas is in thermal (not Knudsen) regime
and
the wall is an excellent thermal conductor (temperature is constant inside the wall).

Each compartment containing perfect gas is described with fundamental relation (\ref{eq:fundamentalRelationGasGravity}).
Since we allow for particle exchange, the energy balance (\ref{eq:energyBalanceGravConsMass}), has to account for the resulting changes in energy $\frac{\partial U}{\partial N} \dd N = \mu^{*} \dd N$.
We do not elucidate this term and treat it formally because it will cancel out due to further derivation.
The amended energy differential is 
\begin{widetext}
 \begin{align}
 \dd U 
 = & T^{*} \dd S^{*} 
 + \frac{N gL}{2} \dd M^{*} 
 - \left(\frac{N k_{B}T^{*}}{V}-\frac{N M^{*}gL}{2V}\right)\frac{V}{L} \dd L 
 + N M g \dd z_1 
 + \mu^{*} \dd N
 \nonumber \\
 = & T^{*} \dd S^{*} + N\frac{gL}{2} \dd M^{*} + \ddbarr W + \mu^{*} \dd N
 = \ddbarr Q  + \ddbarr W  + \mu^{*} \dd N.
 \end{align}
\end{widetext}
We rewrite the energy differential explicitly for both compartments. Compartment 1 is at the bottom, so its $z_1$ stays fixed while $z_1=z_w+L_w$ of the upper compartment 2 is subject to change
\begin{subequations}
 \begin{align}
 \dd U_1 
 = & T^{*}_1 \dd S^{*}_1 + N_1\frac{gL_1}{2} \dd M_1^{*} - \left(\frac{N_1 k_{B}T_1^{*}}{V_1}-\frac{N_1 M_1^{*}gL_1}{2 V_1}\right)\frac{V_1}{z_w - z_1} \dd z_w + \mu_1^{*} \dd N  
 \\
 \dd U_2 
 = & T^{*}_2 \dd S^{*}_2 + N_2\frac{gL_2}{2} \dd M_2^{*} + \left(\frac{N_2 k_{B}T_2^{*}}{V_2}-\frac{N_2 M_2^{*}g L_2}{2 V_2}\right)\frac{V_1}{z_2 - z_w - L_w} \dd z_w + N_2 M g \dd z_w - \mu_2^{*} \dd N.
 \end{align}
\end{subequations}
Here, we set $\dd L = \dd z_w $ and $\dd N = \dd N_1 = - \dd N_2$, which results from the direction of motion of the platform and flux of the gas.

To write the total energy balance in the process of wall motion,
we need to calculate the energetic cost of a passage of particle batch $\dd N$ from compartment 1 to compartment 2 through the porous wall.
One way to do it is to use the metaphor of the external being, just like Maxwell's demon.
We calculate the energy the demon has to spend to displace $\dd N$ particles from the bottom to the top compartment. 
The passage starts when $\dd N$ particles leave compartment 1, which generates energy gain in the demon's account
\begin{equation}
 \ddbarr E_{b1w} =  - \mu_1^{*} \dd N
\end{equation} 
Inserting $\dd N$ particles into the pore end at temperature $T(z_w)$ requires, in addition to the volumetric work, the thermal energy
\begin{equation}
 \ddbarr E_{w1p} = - \frac{3}{2} k_B T(z_w) \dd N,
\end{equation}
which demon has to subtract from the account.
Transition through the region border is accompanied by the volumetric work performed on both sides of the border.
When batch $dN$ leaves compartment 1, the remaining gas has to fill the space. Inside the pore, space for the incoming batch has to be accommodated. Both processes, expansion and compression, happen under the same temperature and pressure, which is continuous over the border.
Therefore, having the same magnitude with opposite signs, they cancel out in the demon's account.
Next, the batch is pushed through the pore, and during that process, it changes the potential energy
\begin{equation}
 \ddbarr E_{w,\textrm{pot}} = - M g L_w \dd N
\end{equation}
and undergoes decompression performing work
\begin{equation}
 \ddbarr W_{w} =  - \left( \int_{z_w}^{z_w + L_w} p(z) \frac{\partial \left(\frac{V}{N}(z) \right)}{\partial z} \dd z \right) \dd N.
\end{equation}
Both quantities need to be supplied by the demon and thus subtracted from the account.
Once the end of the pore is reached, thermal energy
\begin{equation}
 \ddbarr E_{w2p} = \frac{3}{2} k_B T(z_w+L_w) \dd N
\end{equation}
is released and added to the demon's account.
Simultaneously,
the particle needs to equilibrate to the bulk conditions by taking energy
\begin{equation}
 \ddbarr E_{b2w} = \mu_2^{*} \dd N,
\end{equation}
which has to be paid by the demon.
During passage through the region border, again, space has to be emptied and filled on both sides of the pore end at $T(z_w+L_w)$, which does not influence the demon's account.
The final state of the account and thus the energetic cost of the passage results from summing of all listed above is 
\begin{align}
 & \ddbarr E_{\textrm{pass}} 
 = \left( \mu_2^{*} - \mu_1^{*} + M g L_w \right) \dd N 
 + \frac{3}{2} k_B \left( T(z_w + L_w) - T(z_w) \right) \dd N + W_{w} \dd N.
\end{align}
To complete the energy balance, we also need to account for the potential energy change due to the motion of the mass of the wall $M_{w}$ in the gravitational field
\begin{equation}
 \ddbarr E_{\textrm{w,pot}} = M_{w} g \dd z_{w}.
\end{equation}
With the use of $\ddbarr E_{\textrm{pass}}$ and $\ddbarr E_{\textrm{w,pot}}$, we can write the second law of stationary thermodynamics
\begin{align}
  & \dd U_1 + \dd U_2 + \ddbarr E_{\textrm{w,pot}} - \ddbarr Q_1 - \ddbarr Q_2  + \ddbarr E_{\textrm{pass}} 
  \nonumber \\
  & = \left( A \left( p_2(z_w+L_w) - p_1(z_w)\right) + M_{w} g \right) \dd z_w 
  + \left( \frac{3}{2} k_B \left( T(z_w + L_w) - T(z_w) \right) - M g L_w + W_w \right) \dd N  \leq 0.
\end{align}
The first three terms represent the changes in the system. The following four terms represent the fluxes of energies that can be exchanged with the environment, and the last term accounts for additional energetic costs to pass through the obstacle.
We stress that all terms that  include chemical potential $\mu$ are cancelled out.

To calculate $W_{w}$, we use the assumption that the wall is an excellent heat conductor, so that $T(z_w+L_w)=T(z_w)$.
This means that demon had to support the cost of the isothermal expansion
\begin{equation}
 W_w = - k_B T(z_w) \ln \frac{p_2(z_w+L_w)}{p_1(z_w)},
\end{equation}
which we can explicitly insert into the second law of stationary thermodynamics
\begin{align}
  & \dd U_1 + \dd U_2 + \ddbarr E_{\textrm{w,pot}} - \ddbarr Q_1 - \ddbarr Q_2 + \ddbarr E_{\textrm{pass}}
  \nonumber \\
  & = \left( A \left( p_2(z_w+L_w) - p_1(z_w) \right) 
  + M_{w} g \right) \dd z_w 
  - k_B T(z_w) \left( \ln \frac{p_2(z_w+L_w)}{p_1(z_w)} 
  + \frac{M g L_w}{k_B T(z_w)} \right) \dd N  \leq 0.
\end{align}
Out of the first bracket, one reads that the wall will move until pressure acting at the top of the bottom compartment (1) matches pressure at the bottom of the top compartment (2) plus the weight of the wall divided by its surface area
\begin{equation}
 p_1(z_w) - p_2(z_w+L_w)  = \frac{M_{w} g}{A}.
\end{equation}
This is the same as saying that the wall will move until the force necessary to expand the bottom compartment matches the force necessary to compress the top compartment plus to lift the weight of the top compartment ($p_2(z_w+L_w) = p_2(z_2) + \frac{N_2 M g}{A}$ see (\ref{eq:eqstate pz0})) plus lift the weight of the wall.
The second bracket informs about the conditions for the gas flow through the pores. The stationary state is reached when 
\begin{equation}
 p_2(z_w+L_w) = p_1(z_w) e^{-\frac{M g L_w}{k_B T(z_w)}}.
\end{equation}
This means that pressures at both sides of the wall satisfy the jump given by the hydrostatic pressure drop inside the column of gas in the wall.
When expressions in  the first and second brackets vanish, we find the condition for the position of the cylinder inside the whole column
\begin{equation} \label{eq:result2}
 p_1(z_w) = \frac{M_w g}{A \left( 1 - e^{-\frac{M g L_w}{k_B T(z_w)}} \right) } .
\end{equation}
The above equation, for the stationary state, binds the mass of the wall, its thickness, temperature and height above the ground through the pressure at $p_1(z_w)$. 
We see that the pressure needed to squeeze an appropriate amount of gas into the pore depends on the wall's temperature. 
A warm wall requires more of it.

Finally, we expand the fraction in (\ref{eq:result2}) assuming that $\frac{M g L_w}{k_B T(z_w)} \ll 1$ and substitute $p_1(z_w)=\rho_1(z_w) \frac{k_B T(z_w)}{M}$ to find
\begin{equation}
 p_1(z_w) = \rho_1(z_w) \frac{k_B T(z_w)}{M} 
 = \frac{M_w g}{A \left( 1 - e^{-\frac{M g L_w}{k_B T(z_w)}} \right) }  
 = \frac{M_w g}{A} \frac{k_B T(z_w)}{M g L_w} + \ldots
\end{equation}
and the Archimedes principle emerges
\begin{equation} \label{eq:archimedes}
 \rho_1(z_w) L_w A = M_w.
\end{equation}
It states that a body submerged in water will rest exactly where the weight of the displaced fluid is equal to that of the submerged body.
Here, equivalently, punching another pore through the wall will not make it move.

\section{Volumetric heating of the ideal gas separated by the adiabatic wall} \label{sec:4}

{\bf Volumetric heating.} 
We first consider a system without an internal wall in the geometry similar to the one in section~\ref{sec:2}.
The additional energy supplied uniformly throughout the volume is $\lambda$ per unit of time and unit of volume.
It extends the set of parameters controlling the non-equilibrium steady state of this system to $(T_1, T_2,\lambda, V, N)$.
In the steady state, the pressure $p$ and, consequently, the energy density $\epsilon = (f/2) p$ are constant, similarly to the systems discussed in section ~\ref{sec:2}.
For volumetric heating,  the outflow of the heat is balanced by the absorption of heat throughout the system. 
Therefore,  the temperature profile is obtained from the following local energy continuity equation
\begin{equation} 
- \kappa \nabla^{2}T(\vec{r}) = \lambda,   
\label{eq:4.0}
\end{equation}
which assumes the Fourier's law of heat conduction.
The coefficient $\kappa$ is the thermal conductivity, which we assume to be temperature-independent.
We apply the mapping of a non-equilibrium steady-state system to a homogeneous equilibrium system.
Solving Eq.~(\ref{eq:4.0}) for the temperature profile with the boundary conditions $T(z=0)=T_1$ and $T(z=L)=T_2$ we find
\begin{equation}\label{eq:4.1}
T(z)=-\frac{\lambda }{2\kappa}z^2+\left(\frac{T_2-T_1}{L}+\frac{\lambda L}{2\kappa}\right)z +T_1
\end{equation}
  After mapping procedure given by  Eqs ~(\ref{eq:2.1}), we obtain the effective temperature
\begin{equation}\label{eq:4.2}
 T^*=T_1 \mathcal{F}\left(\frac{T_2}{T_1},\frac{\lambda L^2}{\kappa T_1}\right),
\end{equation}
where $\mathcal{F}(u,w)$ is  a dimensionless function given by
\begin{equation}\label{eq:4.3}
\mathcal{F}(u,w)=\frac{\sqrt{(u-1+w/2)^2+2w}}{2\left(\tanh^{-1}{\frac{u-1+w/2}{\sqrt{(u-1+w/2)^2+2w}}}-\tanh^{-1}{\frac{u-1-w/2}{\sqrt{(u-1+w/2)^2+2w}}}\right)}.
 \end{equation}
In the limit $\lambda\to 0$, the effective temperature  $T^*$ reduces to the value given by the equation ~(\ref{eq:2.2a}) for an ideal gas in the heat flux induced by the temperature gradient $T_2-T_1$. 
(This can be seen by using the definition of inverse hyperbolic tangent in terms of logarithms.)
On the other hand, if the temperatures on both outer walls are equal (i.e. $T_2=T_1$), the system reduces to that described in our previous article~\cite{Zhang2021}.
The entropy $S^*(U,V,N)$ is given by Eq.~(\ref{eq:2.11}) with 
$U(S^*,V,N)=(f/2)Nk_BT^*=(f/2)pV$.
For fixed number of molecules $N$, the energy change is determined by the control parameters $(T_1, T_2,\lambda, V)$ through Eq.~(\ref{eq:4.2}).
For an incremental and slow change between steady states that does not disturb the pressure uniformity in the system, we have~\cite{Makuch2022}
\begin{equation} \label{eq:4.4}
 dU=\mkern3mu\mathchar'26\mkern-12mu dQ-pdV \qquad  \mathrm{and} \qquad \mkern3mu\mathchar'26\mkern-12mu dQ=T^*dS^*
\end{equation}

\begin{figure}[htb!]
 \includegraphics[width=.6\textwidth]{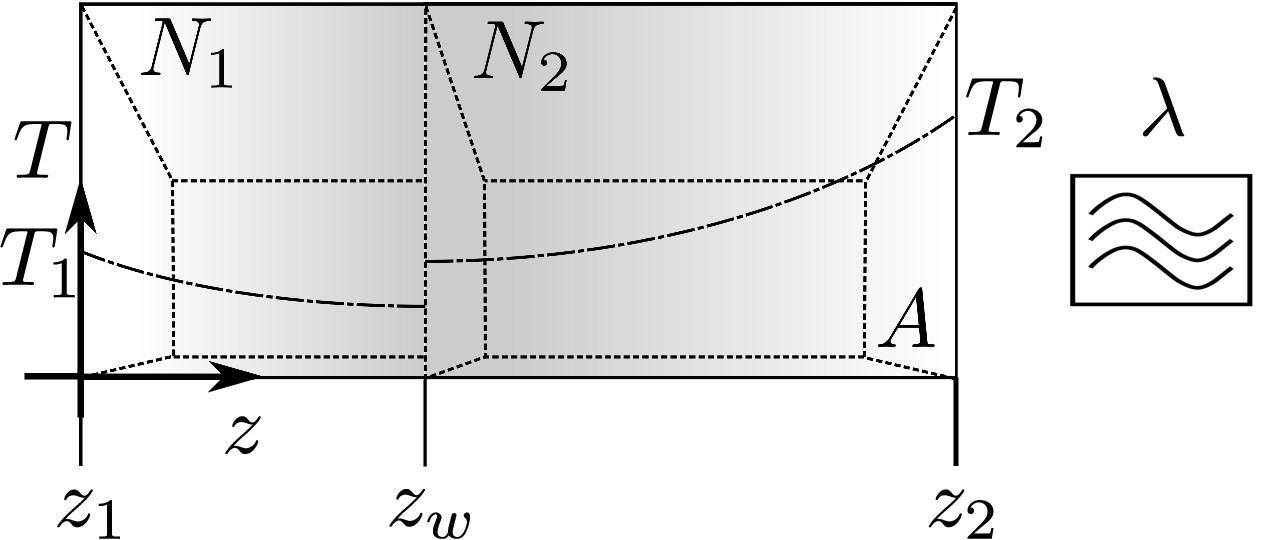}
 \caption[option]{Schematic of a gas confined between two parallel walls of large area $\mathrm{A}$ subjected to
a homogeneous external energy input of density $\lambda$.
The external wall placed at $z_1=0$ is maintained at a temperature $T_1$, while the external wall placed at $z_2=L$ is maintained at a temperature $T_2 >T_1$.
The mobile, thin internal wall is adiabatic, resulting in a jump in the temperature profile.
 }
\label{fig:fig3}
\end{figure}

{\bf Movable wall} We introduce a constraint into the system in the form of a movable adiabatic wall parallel to the bounding walls located at $z = z_w$ (see Fig.~\ref{fig:fig3}).
Under this constraint, the temperature profile has an additional boundary condition, i.e., $\partial T(z)/\partial z =0$ at the internal wall.
The temperature profiles $T^{(1)}(z)$ and $T^{(2)}(z)$ for subsystems   (1) and (2)  with $N_1$ and $N_2$ molecules, respectively are
\begin{equation}\label{eq:4.4}
\begin{aligned}
 &T^{(1)}(z)=-\frac{\lambda }{2\kappa}\left(z-z_w\right)^2+\frac{\lambda }{2\kappa} z_w^2+T_1 \\
 &T^{(2)}(z)=-\frac{\lambda }{2\kappa}\left(z-z_w\right)^2+\frac{\lambda }{2\kappa} (L-z_w)^2+T_2
\end{aligned}
\end{equation}
The mapping procedure gives 
\begin{equation}\label{eq:4.5}
\begin{aligned}
 &T_1^{*}=T_1\mathcal{G}\left(\frac{\lambda z_w^2}{\kappa T_1}\right)\\
 &T_2^{*}=T_2\mathcal{G}\left(\frac{\lambda (L-z_w)^2}{\kappa T_2}\right),
\end{aligned}
\end{equation}
where 
\begin{equation}\label{eq:4.6}
 \mathcal{G}(w) = \frac{\sqrt{w(w+2)}}{2\tanh^{-1}{\sqrt{\frac{w}{w+2}}}}.
\end{equation}
Thus, the energy of the subsystems of $N_1$ and $N_2$ particles under the constraint is
\begin{equation}\label{eq:4.7}
 U=U_{1}+U_2 = \dfrac{f}{2}N_{1}k_{B}T_1 \mathcal{G}\left(\frac{\lambda z_w^2}{\kappa T_1}\right) +\dfrac{f}{2}N_{2}k_{B}T_2 \mathcal{G}\left(\frac{\lambda (L-z_w)^2}{\kappa T_2}\right)
\end{equation}
with $N_1+N_2=N$. 

Our  second  law of non-equilibrium  thermodynamics states 
\begin{equation} \label{eq:4.8}
 \dd U_1 + \dd U_2 - \mkern3mu\mathchar'26\mkern-12mu dQ_1 - \mkern3mu\mathchar'26\mkern-12mu dQ_1 = -\left(p_1 - p_2   \right)  \dd V_1 \leq 0 \; ,
\end{equation}
therefore, the condition for the whole system to reach a steady state is that the pressures exerted by each subsystem are equal. 
This condition is equivalent to  
\begin{equation}\label{eq:4.9}
 \frac{1}{z_w}\mathcal{G}\left(\lambda^* \tilde z_w^2\right) = r\dfrac{N_2}{N_1}\frac{1}{1-\tilde z_w} \mathcal{G}\left(\frac{\lambda^*}{r}(1-\tilde z_w)^2\right),
\end{equation}
where $\tilde z_w =z_w/L$, $r$ is the temperature ratio $r=T_2/T_1$, and $\lambda^* =\lambda/(\kappa T_1)$.
At equilibrium, i.e., for $r=1$ and  $\lambda^*=0$, the internal wall is located precisely in the middle of the system $z_w = 0$.
For $\lambda^*=0$ but $r>1$, the position of the internal wall is given by Eq.~(\ref{eq:2.21}).
\begin{figure}[htb!]
 \includegraphics[width=0.5\textwidth]{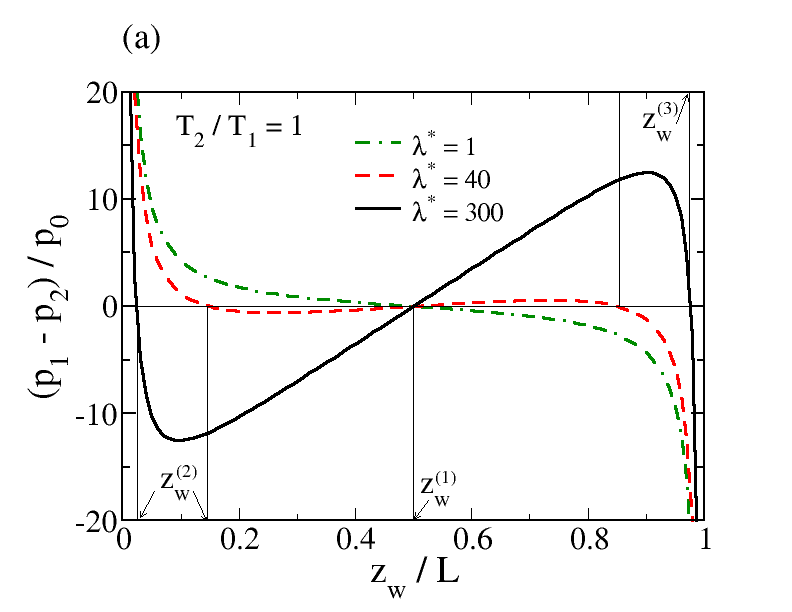}\hspace{-0.01\textwidth}
 \includegraphics[width=0.5\textwidth]{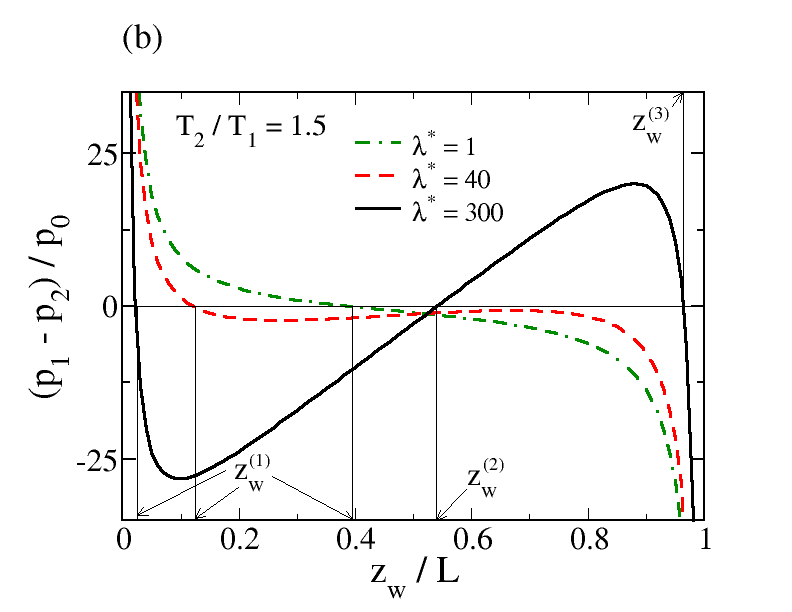}\hspace{-0.01\textwidth}
 \caption[option]{The difference between pressures in two subsystems  normalized with equilibrium pressure $p_{0}=Nk_bT_1/V$, where $T_1$ is the temperature of the external wall located at $z_1=0$
 as a function of $z_w$ for three values of the $\lambda$ and 
 (a) equal temperatures at both external walls  $r=T_2/T_1=1$;
 (b) different temperatures at both external walls  $r=T_2/T_1=1.5$.
 The vertical lines mark the position of the steady states.  In (a)  it is $z^{(1)}_w$ for $\lambda^*=1$ and $z^{(1)}_w,z^{(2)}_w$ and $z^{(3)}_w$ for $\lambda^*=40$ and $300$.
In (b) it is $z^{(1)}_w$ for  $\lambda^*=1$ and  $40$, and $z^{(1)}_w,z^{(2)}_w$ and $z^{(3)}_w$ for $\lambda^*=300$. 
 $\lambda^* = \lambda L^2/(\kappa T_1)$}
\label{fig:mov}
\end{figure}

Analysis of equations~(\ref{eq:4.9}) shows that the number of solutions for $z_w$ varies from one to three depending on the values of $ \lambda^* $ and $r$.
A typical course of the variability of the function $p_1(z_w)-p_2(z_w)$ is shown in Fig.~\ref{fig:mov}  compared to the case of no temperature gradient ($r=1$).
For $r=1$ and small values of $\lambda^*$ the curve $p_1(z_w)-p_2(z_w)$ is a monotonically decreasing function with a single zero-crossing point at the midpoint of the system. The local stability analysis shows that this is a stable position of the internal wall.
Upon increasing the volumetric heating, the function develops a minimum and a maximum but it remains symmetrical with respect to $z_w=1/2$ with three zero-crossing points at $z^{(1)}_w=0$, $z^{(2)}_w=z_w^*$, and $z^{(3)}_w=L-z_w^*$ as illustrated in Fig.~\ref{fig:mov}(a). Among these three solutions,
further analysis shows that locally, both  $z^{(2)}_w$ and  $z^{(3)}_w$ are stable, whereas $z^{(1)}_w$ is unstable. Moreover, the work on changing the position of the wall from $z_w^*$ to $L-z_w^*$
\begin{equation}\label{eq:5.0}
 W_{z^{(2)}_w->z^{(3)}_w}=-\int_{z^{(2)}_w}^{z^{(3)}_w}\left(p_1(z)-p_2(z)\right)dz
\end{equation}
is equal to zero $W_{z^{(2)}_w->z^{(3)}_w}=0$. This, in turn, means that these two stable steady states coexist.

In the case of a non-zero temperature gradient, the symmetry is broken, and the situation changes qualitatively. The single solution which exists for smaller values of $\lambda^*$ is shifted away from the midpoint towards the lower temperature wall, as might be expected.
The difference $p_1(z)-p_2(z)$ becomes non-monotonic as $\lambda^*$ increases, just as it does in the absence of a temperature gradient. 
Initially,  the extremes of the function develop below the $y=0$ axis. Above the value $\lambda^*_c(r)$ for which the maximum of the function $p_1(z)-p_2(z)$ touches the axis $y=0$, 
three zero crossing points $z^{(1)}_w < z^{(2)}_w < z^{(3)}_w$ appear as shown in Fig.~\ref{fig:mov}(b). They correspond to three non-equilibrium steady states. To evaluate their stability, we assume that the internal wall is pushed away from a specific steady state in both directions.
We see that the pressure difference will push the inner wall toward $z^{(1)}_w$ or $z^{(3)}_w$, but push it away from $z^{(2)}_w$.
Thus, the middle position is locally unstable, whereas both positions close to the external walls are locally stable.
Considering the work needed to move the wall from position $z^{(1)}_w$ to position $z^{(3)}_w$ allows us to determine which steady state is globally stable.
If the work $W_{z^{(1)}_w->z^{(3)}_w}$ done on the system during this process is negative, the final position $z^{(3)}_w$ corresponds to a globally stable steady state.
If this work is positive, the initial position $z^{(1)}_w$ is globally stable.
We calculate this work numerically up to the machine accuracy. We find that for $r>1$, due to the asymmetry, the work is positive for all values of $\lambda$; therefore, $z^{(1)}_w$ is always the globally stable steady-state position of the internal wall. 
Thus, the internal wall is globally stable when moved closer to the colder external wall.

\section{Couette flow of the ideal gas}
\label{sec:5}

\begin{figure}[htb!]
 \includegraphics[width=0.5\textwidth]{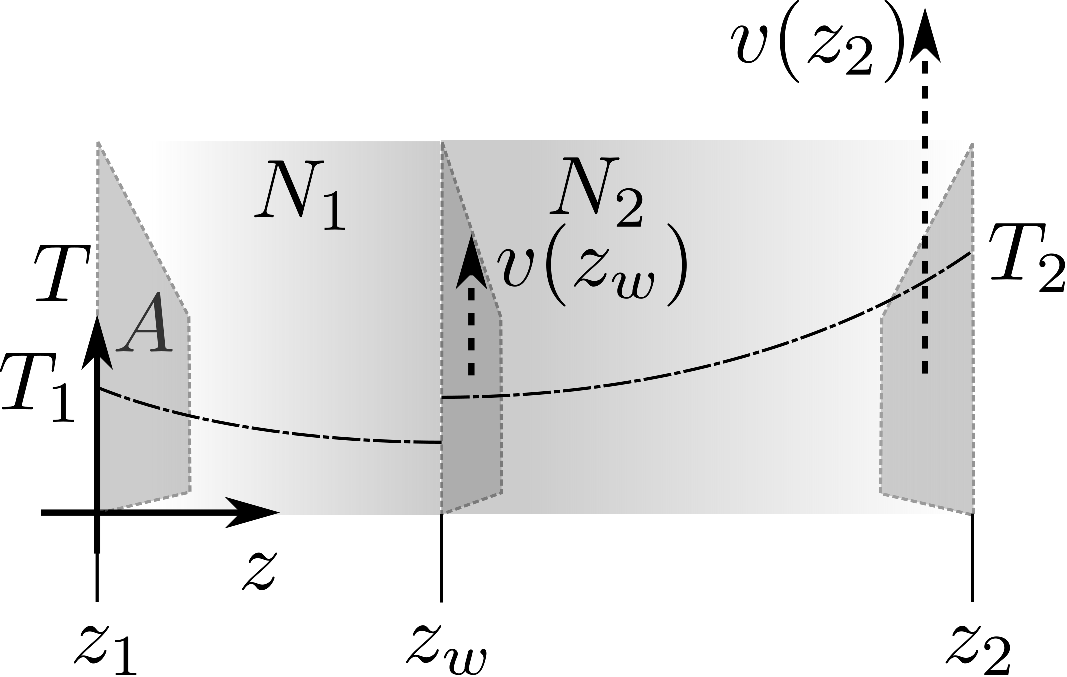}
 \caption[option]{Schematic of a gas confined between two parallel moving walls of large area $\mathrm{A}$.
The external wall placed at $z_1=0$ is fixed and is kept at a temperature $T_1$, while the external wall placed at $z_2=L$ is moving $v_2$ and kept at a temperature $T_2 >T_1$.
The mobile, thin internal wall is adiabatic, resulting in 
the temperature profile jump. }
\label{fig:fig5}
\end{figure}

The last example is the Couette flow of the ideal gas \cite{Karol:JCP2023}. The gas flows between two walls as shown in Fig.~\ref{fig:fig5}. The wall at $z_2=L$ moves at a constant speed $v_2$ along the direction of imposed shear force. The wall at $z_1=0$ is fixed. The velocity profile is linear $v(z)=v_2 z/L$.
Inside the system is an adiabatic massless wall at location $z_w$, which moves at a local speed of the fluid $v_w= v_2 z_w/L$ along the same direction as the upper wall. 
The wall divides the system into two parts,
1 and 2. The location of the wall is the internal constraint in the system. We ask the following question: if we let the wall move along the $z$ direction, what will be the final stationary location of this internal wall? This system differs from all the previous ones in two ways. First of all, the total energy includes internal energy and kinetic energy. The latter was absent in the cases discussed in the previous paragraphs. 
Secondly, the system exchanges the energy with the external world in the form of heat and work. Continuous input of work keeps the constant flow and constant kinetic energy despite dissipation due to shear. In the steady state at a fixed location of the internal wall, the work dissipates as heat inside the system. When we release the internal wall, it starts moving in the $z$ direction. During its motion, the external machine must do extra work 
to overcome extra shear (for the constant velocity of the upper wall) and dissipation. 

The internal wall divides the system into two subsystems: 1,2. From the equations of irreversible thermodynamics \cite{Karol:JCP2023} the change of the internal energy $\dd U_i$ for subsystem $i$ (i=1,2) is given by:
\begin{equation}
	\dd U_i= \ddbarr Q_i + \ddbarr W_i + \ddbarr D_i,\label{eq:first law int en}
\end{equation}
where $\ddbarr Q_i$ is the net heat in subsystem $i$ ($i=1,2$), $\ddbarr D_i$ is the dissipation in subsystem $i$ and $\ddbarr W_i=-p_i \dd V_i$ is the volumetric work done during the transition in subsystem $i$. In this non-equilibrium case, we get $T^*\dd S^*= \ddbarr Q+ \ddbarr D$. Thus, the entropy change is due to both net heat exchange and dissipation due to shear inside a system. $\dd V_i$ is the change of the volume in subsystem $i$. From the equation of irreversible thermodynamics \cite{Karol:JCP2023} it follows that the change of the kinetic energy, $E_{k,i}$ in the $i$ subsystem ($i=1,2$) has the following form:
\begin{equation}
	\dd E_{k,i}=\ddbarr W_{w,i}- \ddbarr D_i,\label{eq:first law kin en}
\end{equation}
The work done by the surface forces $\ddbarr W_{w, i}$ during the motion of the internal wall in the $z$ direction keeps the constant velocity profile in the system despite changes of the shear forces at the walls. We formulate the second law as follows. From the sum of the total energy of the two subsystems, we subtracted the heat and work exchanged with the environment during the processes and demanded that the difference be non-positive:
\begin{equation}
\begin{aligned}
 \dd U_1+\dd U_2+ \dd E_{k,1}&+ \dd E_{k,2} 
 - \ddbarr Q_1- \ddbarr Q_2  \\
 &- \ddbarr W_{w,1} - \ddbarr W_{w,2} \le 0 \label{eq:second law}
\end{aligned}
\end{equation}
From the equation, we get the expected form of the second law:
\begin{equation}
	-(p_1-p_2)\dd V\le 0
\end{equation}
The equality in the equation defines the condition of the stationary position of the wall, given by the equality of pressures $p_1=p_2$ in two subsystems.

\section{Discussion}

The system at equilibrium, which exchanges heat with the environment, satisfies the inequality $dU-dQ\le 0$ at a constant temperature. 
In this contribution, we showed that
the same inequality holds for non-equilibrium states at constant boundary conditions, including boundary temperature. 
This inequality is the second law of non-equilibrium thermodynamics for systems coupled to the environment via heat flux. 

In equilibrium thermodynamics, at constant temperature $T$ and pressure $p$, the second law states that $dU-TdS-pdV \le 0$.
We can write it in a more general form as $dU-dQ-dW\le 0$. 
In the example of the Couette flow, similar inequality sets the direction of spontaneous processes in the system. 
In general, in a system that exchanges energy by heat, $dQ$, and different forms of work $dW_j$, the following inequality should govern spontaneous processes:
\begin{equation}
	dE-dQ-\sum_jdW_j\le 0.
\end{equation}
$E$ is the system's total energy, including internal, potential, and kinetic energy.
The most challenging part of our study was identifying various terms in this equation and calculating net heat and work done during the process. 

We are now in a position to discuss the second law in more detail with all subtleties.
The second law of the equilibrium thermodynamics for the isolated system states that the entropy of the system has the maximum value at equilibrium. However, $S(U,V,N)$ has a well-defined value for fixed $U,V,N$. 
Therefore, we have to compare the entropy of the system in the equilibrium state to the entropy of the system at states that are not accessed by simply changing the state parameters $U,V,N$. 
We introduce an extra parameter $x$, usually in the form of a constraint in the system (like a moveable wall) and state that $S(U,V,N,x)$ is maximized as a function of $x$ for fixed state parameters $U,V,N$ (internal energy, volume and the number of particles respectively). 
Similarly, when the system has a constant temperature, the Helmholtz free energy, $F(T, V, N, x)$, is minimized as a function of $x$ (the variable describing internal constraint) at constant $T, V, N$ (temperature, volume, and number of particles, respectively). 
We treat this extra parameter as arising from the external device, i.e., an external device that performs work on the system by coupling to this internal parameter $x$ and moves the system in a reversible way (i.e. via a sequence of equilibrium states) between states that are not accessible by simply changing $T,V,N$. Let us define the work done in the system by this external device by $dW_z$. In the process of going between the states, the first law of thermodynamics must be obeyed (conservation of energy). Thus, we have $dU=dQ+dW_z$. Now, the second law states that if we move a system from a stable state to a less stable state, the external device will do work on our system i.e. $dW_z>0$. 
On the other hand, if we move from a less stable state to a more stable state, the system will do work on the device and $dW_z\le 0$. 
This last inequality is yet another statement of the second law.
For example, in the case of the movable wall, we can apply an external force to the wall and move the system between states, where the pressures on both sides of the internal wall are different and satisfy the equality $p_1-p_2=F/A$, where A is the area of the wall and $F$ the external force.
Now, we can rewrite the first law of thermodynamics in the following form $dU-dQ=dW_z\le 0$ and thus get $dU-dQ\le 0$.
The last form is more convenient than $dW_z\le 0$, because we don't have to create for each case a new device and introduce new parameters. 
Everything can be calculated from the system's state if we know $dU$ and $dQ$. 
However, the form $dU-dQ$ compares the neighbouring states and is, therefore, local.
In non-equilibrium states, this form is not enough for the prediction of the direction of the spontaneous process. 
In many non-equilibrium situations, we may have many stable local states. For example, in section IV, we described the volumetric heating of a gas.
There, we had three different states that satisfied $dU-dQ=0$.
The question is how to compare them and decide which of them is most stable?
We propose in this paper to calculate the total work done along the path i.e. $\int dW_z$.
If this total work is negative, it means that we moved from a less stable state to a more stable state. 
Thus, the second law would be $\int (dU-dQ)\le 0$. Because work depends on the chosen path, we, in the end, demand that this work be the maximal, i.e. ${\rm max}\int (\dd U- \dd Q)\le 0$. The calculations have to be done at constant boundary conditions. The second law, in simple language, states that the direction of the spontaneous process removes some energy from the system by performing work on the external device that keeps the system in a less stable state. 

This contribution, together with our previous works on the first law of non-equilibrium thermodynamics, constitutes a good starting point to apply the presented here second law of non-equilibrium thermodynamics to systems undergoing chemical reactions in photo-reactors and flow reactors, to Rayleigh-Benard convection, thermoosmosis and finally, lift force in hydrodynamic flows.

\section{Acknowledgements}

P.J.\.{Z}. would like to acknowledge the support of a project that has received funding from the European Union's Horizon 2020 research and
innovation program under the Marie Sklodowska-Curie Grant Agreement
No. 847413 and was a part of an international co-financed project
founded from the program of the Minister of Science and Higher Education entitled ``PMW'' in the years 2020--2024, Agreement No. 5005/H2020-MSCA-COFUND/2019/2.
This research was funded in part  by  the Polish National Science Center (Opus Grant  No.~2022/45/B/ST3/00936).

\bibliographystyle{unsrt} 

\end{document}